\documentclass[baaa]{baaa}

\usepackage[pdftex]{hyperref}
\usepackage{natbib}
\usepackage{helvet,soul}
\usepackage[font=small]{caption}
\usepackage{float}

\usepackage{soul}
\sethlcolor{lime}
\usepackage{lipsum}
\usepackage{graphicx}   
\usepackage{subcaption}  
\setkeys{Gin}{width=\linewidth,height=0.28\textheight,keepaspectratio}

\contriblanguage{1}

\contribtype{1}

\thematicarea{5}

\received{\ldots}
\accepted{\ldots}

\title{HC$_3$N, H$^{13}$CN, and HN$^{13}$C in molecular cores evolving towards star-forming regions}

\titlerunning{Chemistry in molecular cores evolving towards star-forming regions}

\author{
R. D. Taboada\inst{1,2,3,4},
S. Paron\inst{4},
M.E. Ortega\inst{4},
\& H. Saldaño\inst{1}
}

\authorrunning{Taboada R. D. et al.}

\contact{rocio.taboada@exa.unsa.edu.ar}

\institute{Instituto de Investigaciones en Energía No Convencional, CONICET-UNSa, Argentina
\and Departamento de F{\'\i}sica, Facultad de Ciencias Exactas y Naturales, UBA, Argentina
\and Departamento de Matemática,  Facultad de Ciencias Exactas, Universidad Nacional de Salta, Argentina 
\and Instituto de Astronomía y Física del Espacio, CONICET-UBA, Argentina
}

\resumen{Como parte de un trabajo en progreso se presentan resultados de un análisis químico y físico de núcleos moleculares en estadios evolutivos tempranos en lo que respecta a la formación estelar. A partir de datos de archivo del Atacama Large Millimeter Array (ALMA) se investiga una muestra de 37 fuentes, de las cuales se extrajeron espectros en un rango de frecuencias de 330--350 GHz hacia las posiciones centrales de los núcleos moleculares. Se analizaron transiciones de HC$_3$N, H$^{13}$CN, y HN$^{13}$C mediante ajustes gaussianos, obteniendo intensidades máximas, flujos y anchos de línea. Se estimaron las densidades de columna de cada molécula y sus abundancias. Se estudió el comportamiento de dichas abundancias con la temperatura de cada región, observándose correlaciones positivas para  H$^{13}$CN y HN$^{13}$C, y ninguna correlación aparente para el HC$_3$N. Este estudio contribuye a la caracterización de las condiciones iniciales del medio interestelar en fases tempranas de evolución estelar.}

\abstract{As a work in progress, results from a chemical and physical analysis of molecular cores in early evolutionary stages concerning star formation are presented. Using archival data from the Atacama Large Millimeter Array (ALMA), a sample of 37 sources was investigated, from which spectra in the frequency range 330--350 GHz were extracted towards the central positions of the molecular cores. Transitions of HC$_3$N, H$^{13}$CN, and HN$^{13}$C were analysed using Gaussian fits, obtaining peak intensities, fluxes, and line widths. The column densities of each molecule and their abundances were estimated. The behaviour of these abundances with the temperature of the region was studied, observing positive correlations for H$^{13}$CN and HN$^{13}$C, and none for HC$_3$N. This study contributes to the characterisation of the initial conditions of the interstellar medium in early phases of stellar evolution.
}

\keywords{ISM: molecules --- Stars: formation --- Stars: evolution}

\begin{document}

\maketitle
\section{Introduction}
\label{intro}

Molecular cores embedded in clumps can be used as astrochemical laboratories to investigate how complex molecules
are formed in space (e.g., \citealt{jor20,coletta20}). The understanding of the interstellar chemistry allows us to characterise the gas and dust 
condensations where stars form. Indeed, the molecular matter is present at all spatial and temporal scales related to star formation \citep{herbst09,bonfand19,min23}. 

Molecular cores at the early stages of star-forming evolution are of high interest to constrain the early chemistry in such environments. In previous works \citep{sulfur,paron25} some of the authors of this article focused on the analysis of the sulphur chemistry at a sample of 37 cores embedded in some of the most massive infrared-quiet molecular clumps from the Atacama Pathfinder EXperiment (APEX) Telescope Large Area Survey of the Galaxy (ATLASGAL). As a continuation of this astrochemical work, using the same sample of sources, we decided to study other kind of molecules such as HC$_{3}$N, H$^{13}$CN, and HN$^{13}$C.

It is well known that the shortest cyanopolyyne HC$_{3}$N, the cyanoacetylene, is useful to explore gas associated with hot molecular cores 
\citep{bergin96,taniguchi16,duronea19}. In addition it was suggested that the HC$_{3}$N would trace not only the chemistry generated in the envelopes 
of the hot cores but also that related to the shocked gas \citep{hervias19}. 
The hydrogen cyanide (HCN) and isocyanide (HNC), and their isotopes, have a linked chemistry, and differences in the spatial distributions in which they are observed within a cloud can reflect the gas chemical conditions and the evolution of the star-forming regions \citep{shcilke92}. For this reason, they have often been used as probes for the chemical evolution of the gas in both dense and starless
cores and in star-forming regions (\citealt{lefloch2021} and references therein). Additionally, it was proposed that these isomeres can be used to estimate temperature through intensity ratios \citep{hacar20,nai23,nai24}. However, as pointed out by \citet{nai24}, the analysis of HCN and HNC may present some difficulties due to high-optical depth effects that the lines can suffer, which makes the analysis of the optically thinner isotopes more accurate. 

In conclusion, investigating these quite simple molecular species at the earliest stages of star formation is important to shed light on the subsequent chemical processes that will take place, as they should be involved in the chemistry that yields more complex molecules.

\section{Data and source sample}
\label{sectdata}

The data set was retrieved from the Atacama Large Millimeter Array (ALMA) Science Archive\footnote{http://almascience.eso.org/aq/} (project 2017.1.00914; PI: Csengeri). This observing project consisted of observations in Band 7 (frequency range  333.3–349.1 GHz) towards massive infrared-quiet clumps in the inner Galaxy, i.e. sources at early stages of star formation. The spectral and angular resolution are 1.1 MHz and 3.5$^{\prime\prime}$, respectively.

From a sample of 37 molecular clumps, we analysed the main cores embedded within them. The catalogued clumps, the cores coordinates, their local standard of rest velocities, and distances are presented in Table\,A.1 of Appendix\,1 in \citet{sulfur} (we do not present it here due to space constraints). Table\,\ref{transitions} presents the analysed molecular lines towards the centre of each core, including the rotational transition and the rest frequency ($\rm \nu_{rest}$). 

\begin{table}[h]
\centering
\small
\caption{Analysed molecular lines.}
\label{transitions}
\begin{tabular}{llc}
\hline
\hline
Molecule & Transition & $\rm \nu_{rest}$ (GHz)   \\
\hline
 & &   \\[-1.8ex]
HC$_{3}$N v=0    & J=37--36  & 336.5200      \\
HC$_{3}$N v=0    & J=38--37  & 345.6090     \\
H$^{13}$CN v=0   & J=4--3    & 345.3397     \\
HN$^{13}$C       & J=4--3    & 348.3402     \\
\hline
 & &   \\[-1.8ex]
\end{tabular}
\end{table}

\section{Results}\label{res}

\begin{figure}
  \centering
   \includegraphics[width=\linewidth]{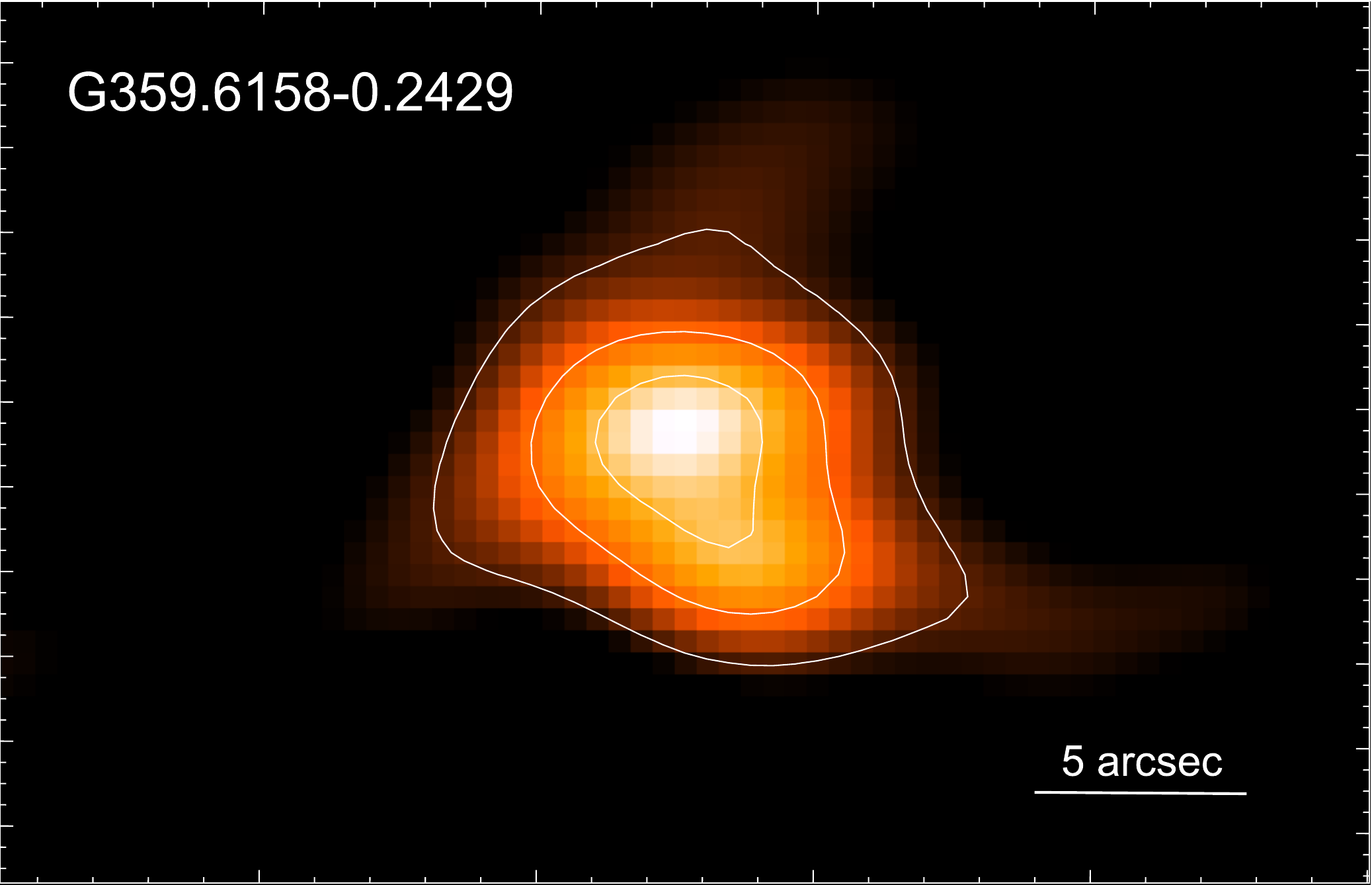}
  \caption{As an example, we show the ALMA  continuum emission at 0.8 mm from the core embedded in the ATLASGAL source G359.6158$-$0.2429. The contour levels are 0.7, 1.2, and 1.7 Jy beam$^{-1}$. From a beam area centred at the peak of the core, the spectra presented in Fig.\,\ref{fig:espectros} were extracted. The rms noise level is 0.02 Jy beam$^{-1}$. }
  \label{core}
\end{figure}

\begin{figure}[h!]
  \centering
  \includegraphics[width=\linewidth]{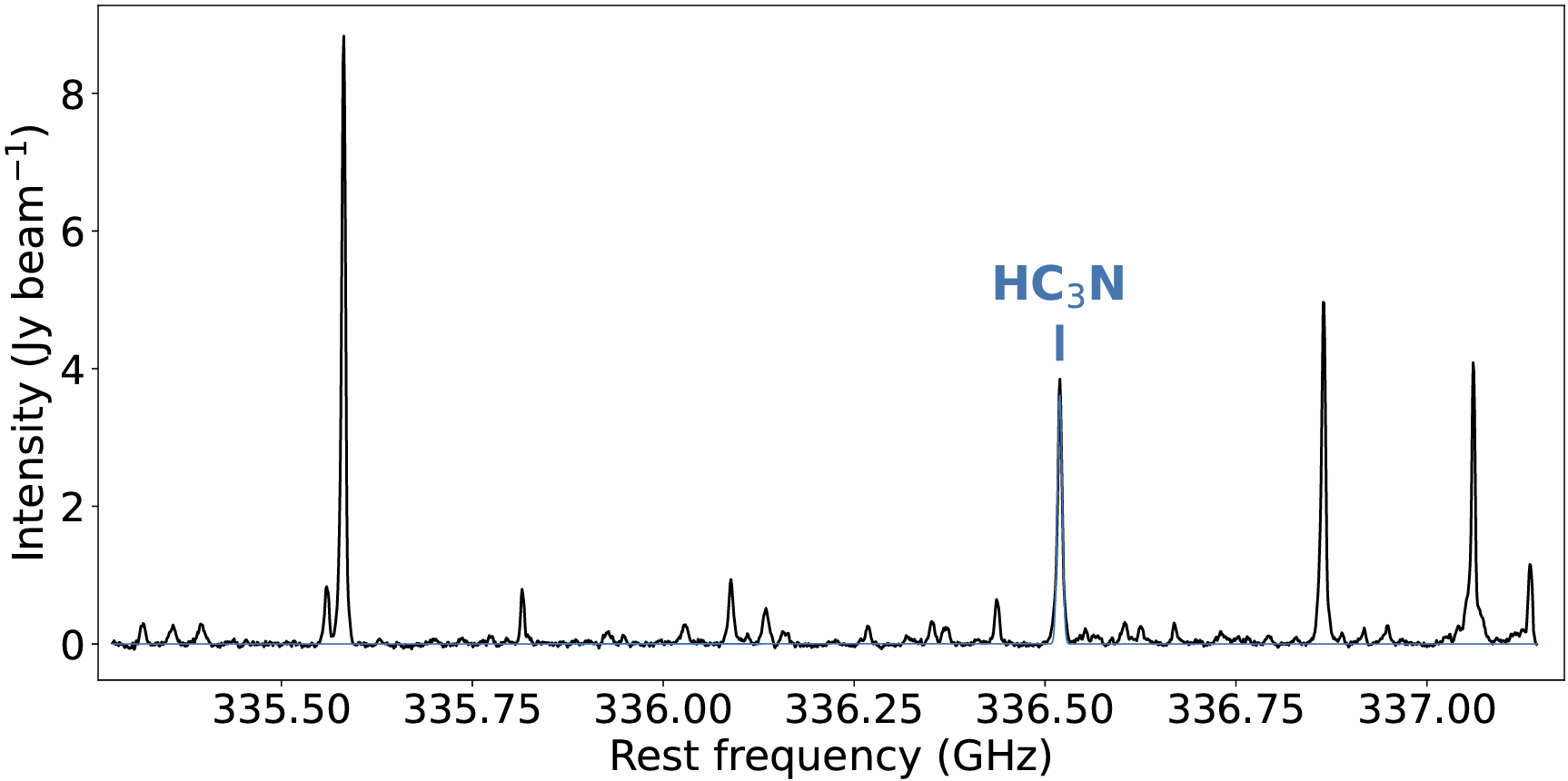}\\[2mm]
  \includegraphics[width=\linewidth]{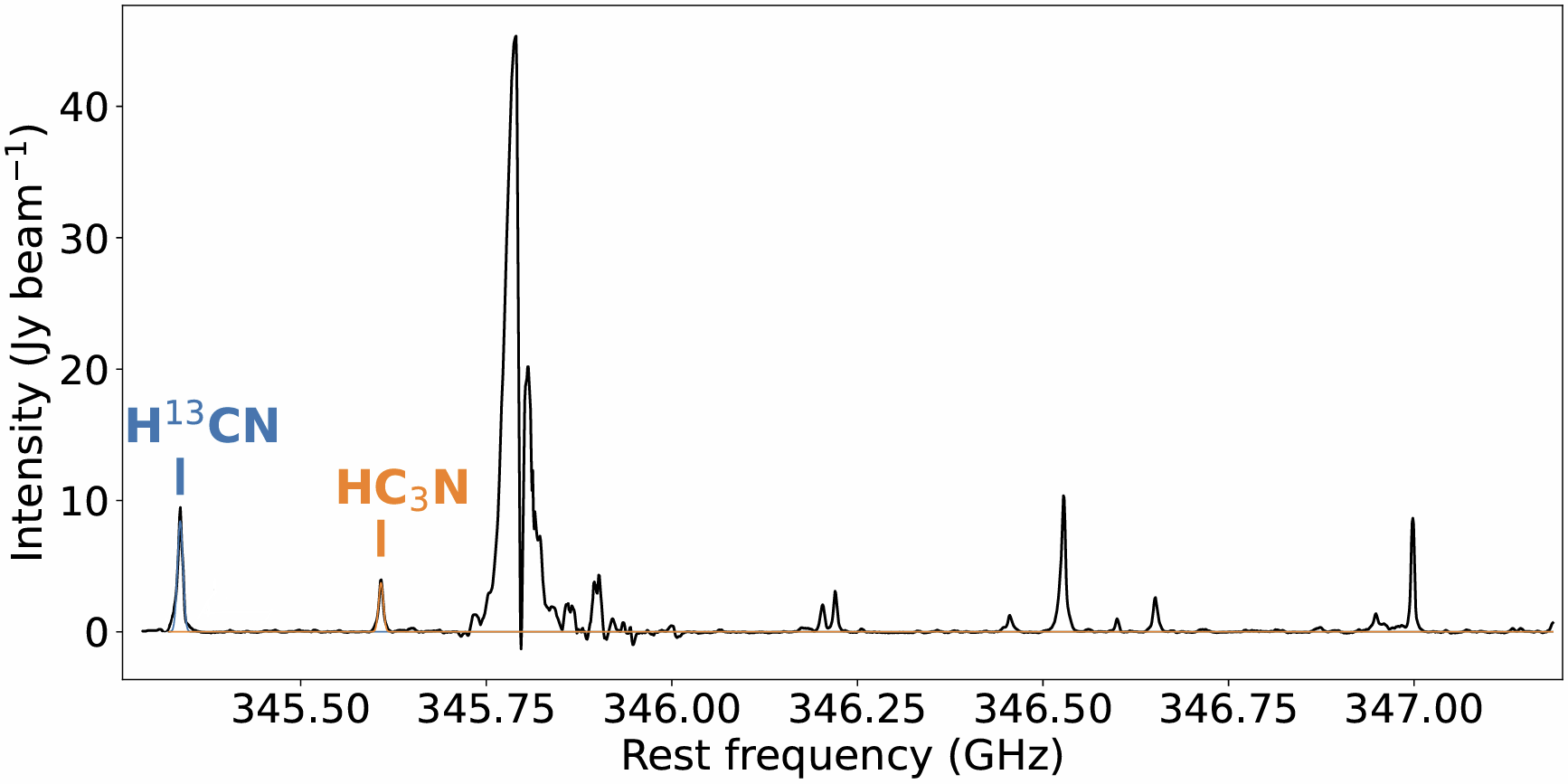}\\[2mm]
  \includegraphics[width=\linewidth]{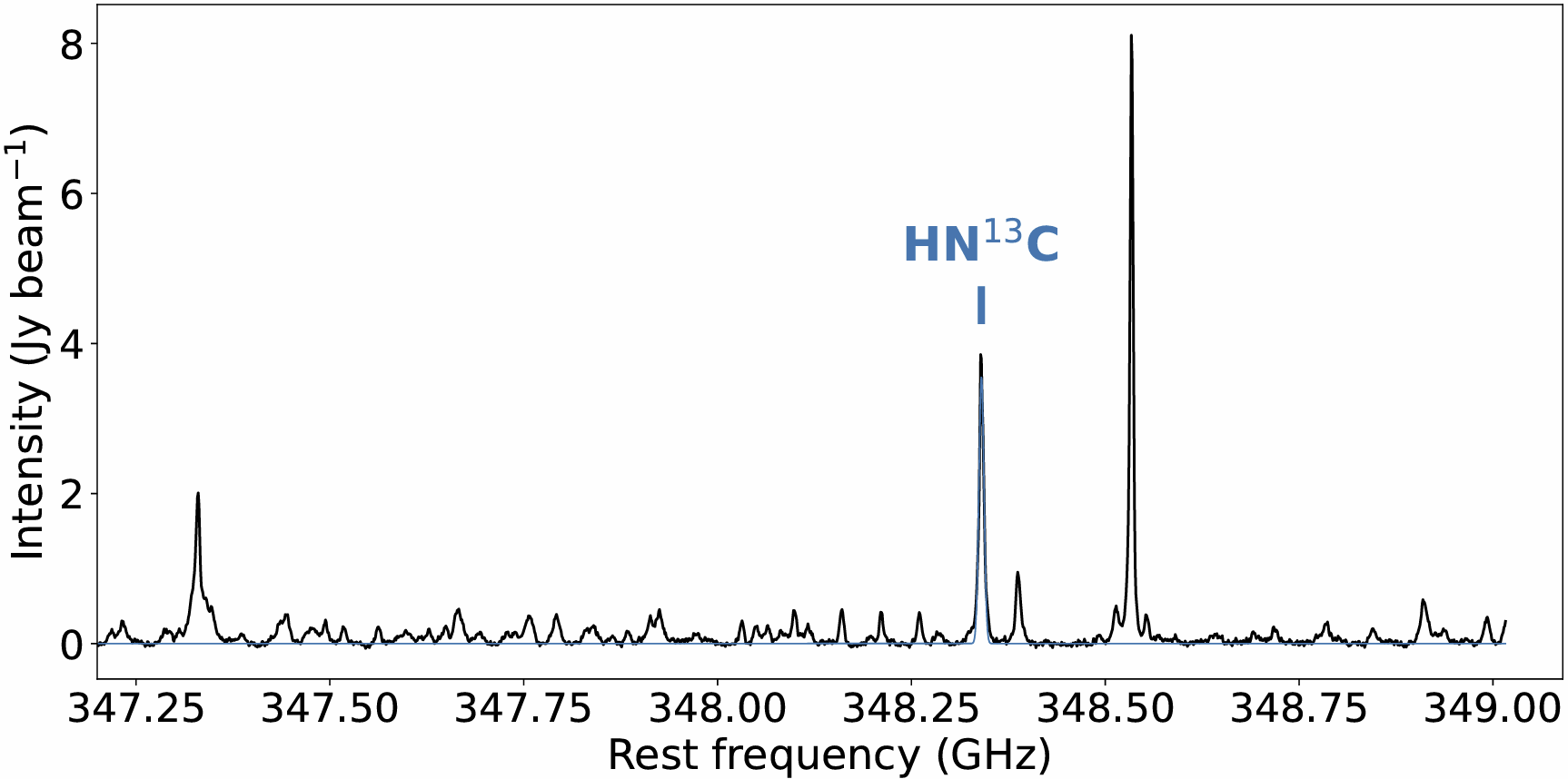}
  \caption{Spectra obtained from the core presented in Fig.\,\ref{core} in which the transitions of HC$_3$N (detected in two lines), H$^{13}$CN, and HN$^{13}$C are identified. Gaussian fittings to the selected lines are displayed in colours. These spectra are shown as an example of the spectral analysis performed in the whole sample of 37 cores. All identified lines have signal to noice (S/N) ratios above 100. }
  \label{fig:espectros}
\end{figure}

For each of the 37 molecular cores identified in the clump sample, the lines of HC$_3$N, H$^{13}$CN, and HN$^{13}$C presented in Table\,\ref{transitions} were identified in the spectra obtained from a beam area at the peak position of the cores continuum emission. In Fig.\,\ref{core} we present as an example the continuum emission of the core embedded in the ATLASGAL source G349.6158$-$0.2429. The spectra presented in Fig.\,\ref{fig:espectros} were extracted from its peak position.

The spectra presented in Fig.\,\ref{fig:espectros} as an example shows Gaussian fittings performed to the selected lines. This procedure was done for the complete sample, obtaining the peak intensity, the full width half maximum (FWHM) line width ($\rm \Delta v_{\rm FWHM}$), and the integrated line intensity or flux (W) for each molecular line at each core. 
It is worth noting that, as shown in Fig.\,\ref{fig:espectros}, the spectra do not exhibit a large number of molecular lines. This is consistent with the nature of the sources, which are early-stage and generally cold molecular cores.This issue simplifies line identifications and fittings. Unfortunately many spectra in the 345--347 GHz range are cut off near 345.3 GHz leaving out the H$^{13}$CN  J=4--3 emission. We were only able to analyse the H$^{13}$CN J=4--3 line in five cases. Although there are few points, at least they are useful for exploring some possible trend in the case of this molecular species. 

Once the Gaussian parameters were obtained from the observed lines, we estimated the molecular abundances. 

The population in the upper energy level of a given molecular line was obtained from:

\begin{equation}
{\rm N_u} = 2375\times10^{6}~{\rm W}~{\rm A_{ul}^{-1}}~{\rm \Theta_{\rm area}^{-1}}~,
\label{eq:Nu}
\end{equation}

\noindent where W is the integrated line intensity or flux (in Jy beam$^{-1}$ km s$^{-1}$), $\rm A_{ul}$ is the Einstein coefficient of the transition (in s$^{-1}$), and $\Theta_{\rm area}$ is the beam area (in arcsec$^2$).
Assuming local thermodynamic equilibrium (LTE) conditions, the column density (N) is obtained from:

\begin{equation}
{\rm N = \frac{N_u Q(T_{\rm ex})}{g_u}\exp\left(\frac{E_u}{T_{\rm ex}}\right)},
\label{eq:Ntot}
\end{equation}

\noindent where Q(T$_{\rm ex}$) is the partition function at the excitation temperature (T$_{\rm ex}$), $\rm g_u$ is the statistical weight of the upper level of the line, and $\rm E_u$ its energy. These parameters were taken from the Splatalogue database\footnote{https://splatalogue.online/\#/advanced} from each analysed molecular line. The excitation temperature of the analysed transitions was assumed to be equal to the kinetic temperature ($\rm T_k$) of the gas in each core. The values of $\rm T_{k}$ were obtained in \citet{sulfur} from the rotational diagram method applied to CH$_3$OH transitions for each core. In all cases, following the Splatalogue database, the value of the partition function Q was selected according to the core temperature.  

Then, having the column densities for each species (N(molecule)), the relative abundances with respect to H$_2$ were derived as:

\begin{equation}
\rm X(molecule) = \frac{N(molecule)}{N(\mathrm{H}_2)} ,
\label{eq:abundancia}
\end{equation}

\noindent where $\rm N(H_2)$, the molecular hydrogen column density, was obtained from the 0.8 mm dust continuum emission (see \citealt{sulfur}).

\begin{figure}[h!]
\centering
\includegraphics[width=8cm]{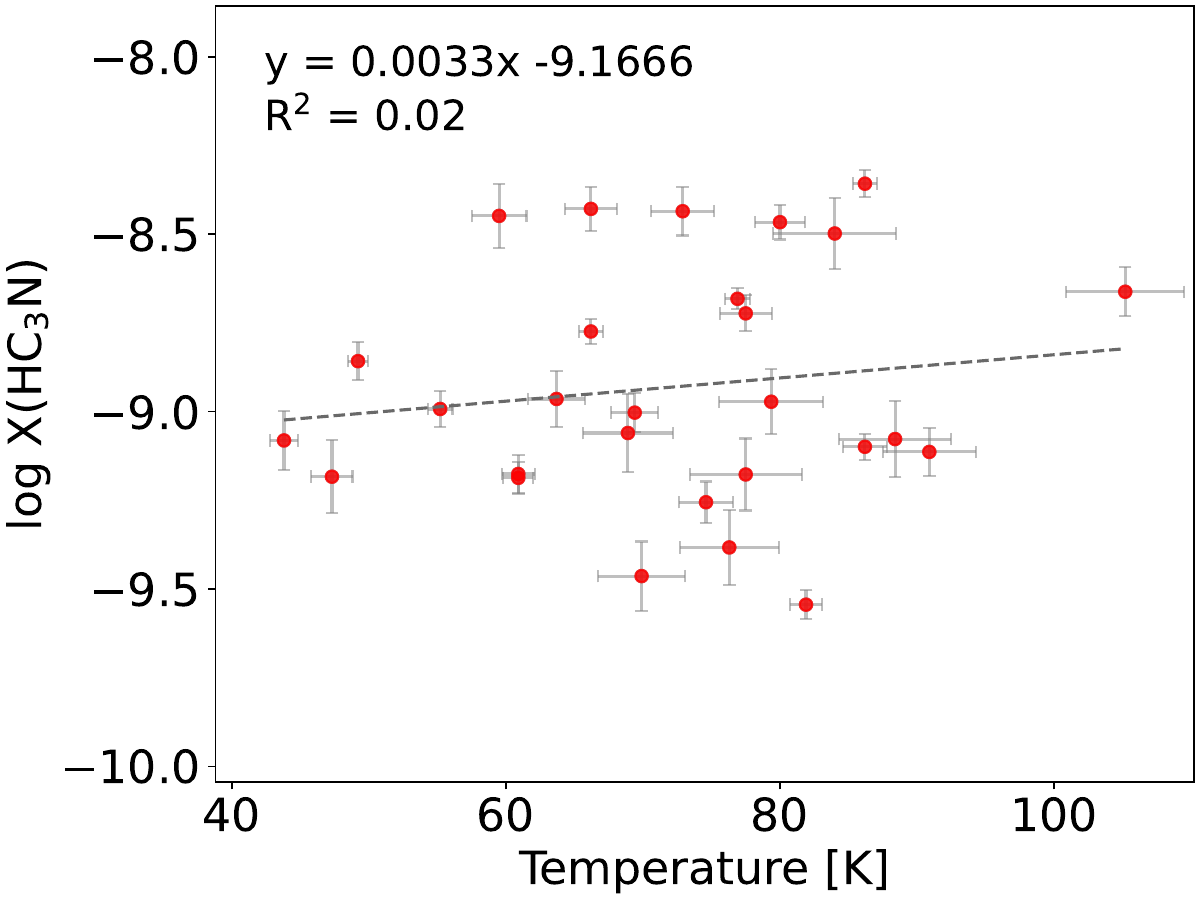}
\includegraphics[width=8cm]{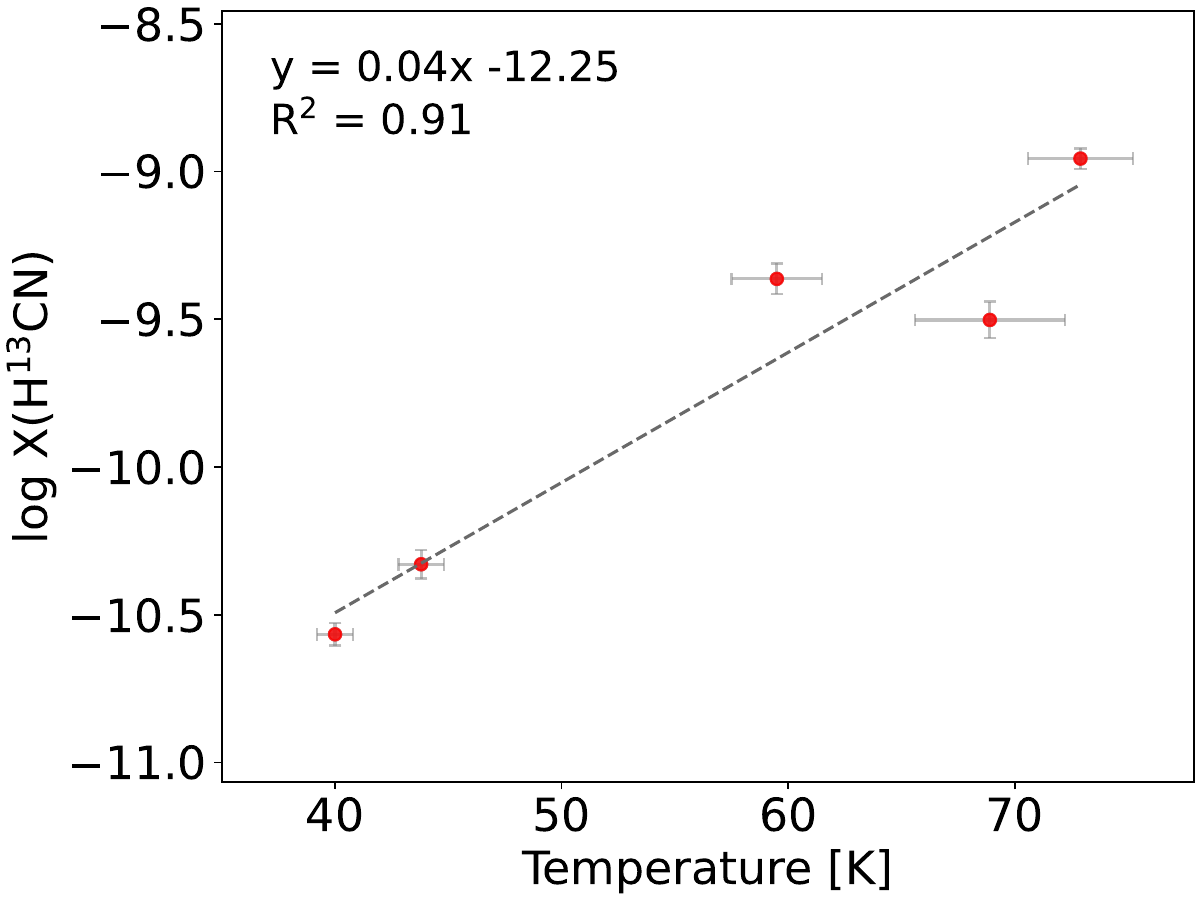}
\includegraphics[width=8cm]{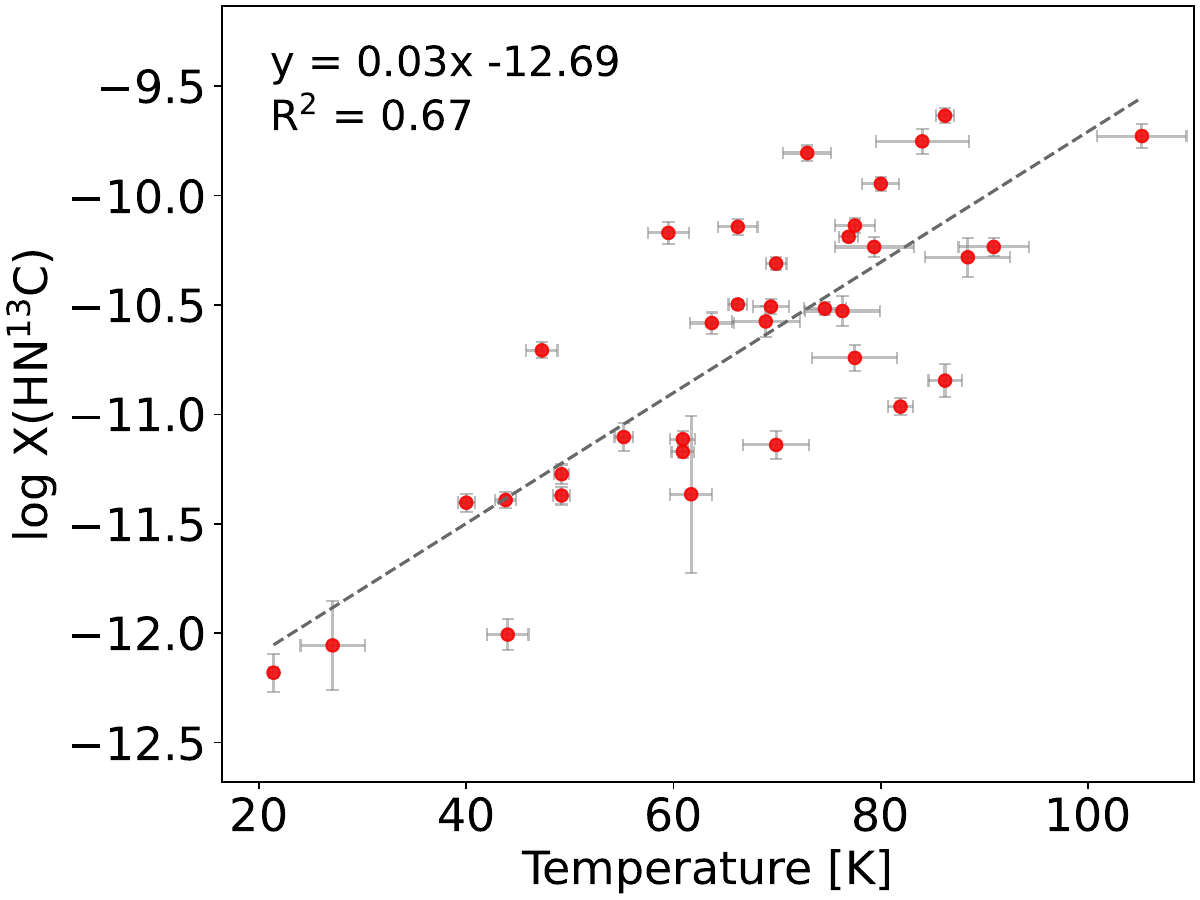}
\caption{Molecular abundances X (in logarithmic scale) vs. kinetic temperature. The lines are the result of linear fits whose results are included in the top left corner of each panel.}
\label{fig:abundancias}
\end{figure}

In order to analyse the behaviour of HC$_3$N, H$^{13}$CN, and HN$^{13}$C with the temperature, in Fig.\,\ref{fig:abundancias} we present their abundances in log scale vs. $\rm T_{k}$. Also the parameters obtained from linear fittings are included.

It is important to note that we are reporting here the detection of two different transitions of HC$_{3}$N(see Table\,\ref{transitions} and Fig.\,\ref{fig:espectros}). The results presented in Fig.\,\ref{fig:abundancias} (upper panel) correspond to the HC$_{3}$N J=37--36 line. The emission of the other line is less frequently detected across the core sample. However, in the case of cores that present emission of both lines, very similar column density values were obtained using either line, supporting the LTE assumption. In such cases,  comparisons between the lines (e.g. intensity ratios) will be presented in a future work aiming to explore the cyanoacetylene excitation conditions in this kind of regions.

\section{Discussion}

Our results presented in Fig.\,\ref{fig:abundancias} show a positive correlation with the temperature in the case of the abundances of HN$^{13}$C and H$^{13}$CN. It should be noted that, in the latter case, the number of available data points is limited, which introduces an additional degree of uncertainty in the trend. The resulting slopes of the linear fittings in the case of X(HN$^{13}$C) and X(H$^{13}$CN) vs. temperature are very similar to those obtained for sulfur-bearing molecules in \citet{sulfur}.  On the other hand, the behaviour of X(HC$_3$N) appears to be almost constant over the whole range of temperatures analysed. This issue represents one of the main findings of this work.

The increase of HN$^{13}$C with kinetic temperature is consistent with scenarios in which the release of HCN and HNC from the icy mantles of dust grains and production of the chemically related species HCNH$^{+}$ take place in warmer environments, thereby enhancing the gas-phase abundance of these species, and likely of their isotopologues \citep{viti2004,long21,gong25}.  
A similar, though less robust, correlation is also observed for H$^{13}$CN, reinforcing the idea that the thermal conditions of the gas favour the presence of this kind of molecules in the gas phase. However, in this case the number of available data points is limited, which reduces the statistical significance. The observed trend should be confirmed with future analyses based on a larger sample of sources with detection of this molecular species.

On the other hand, the abundance of HC$_3$N does not show a clear dependence on temperature, suggesting that its chemistry is not dominated by temperature-sensitive processes at these core evolving stages, or at this temperature range. The chemical models presented by \citet{taniguchi2019cyanopolyyne} predict that during the cold phase of collapse ($\rm T \lesssim 25\ K$), HC$_3$N forms mainly through ion--molecule and neutral--neutral gas-phase reactions that remain efficient even at low temperatures. This may account for the observed stability of HC$_3$N abundance in our sample, which likely precedes the onset of warm-up processes such as warm carbon-chain chemistry and thermal desorption. A fraction of HC$_3$N formed in the cold phase can also freeze onto dust grains and be later released, representing a chemical inheritance from earlier, less evolved stages. 

Following \citet{taniguchi2019cyanopolyyne}, the HC$_3$N abundance measured in our core sample across the reported temperature range could be explained by the production of this molecular species mainly through the gaseous reaction: $\rm C_{2}H + CN \rightarrow HC_{3}N + H$. Partial depletion onto dust grains would compensate for the increase of this molecular species in the gaseous phase. Later, with higher temperatures ($\gtrsim 90$ K), the frozen-out HC$_{3}$N  will be sublimated from such dust grains, enhancing its gaseous abundance \citep{wang2025}. To probe this issue, it would be interesting to study a sample of cores warmer than those presented here.


The observed non-correlation between the HC$_3$N abundance and the temperature contrasts with what has been reported for sulfur-bearing molecules \citep{sulfur}, where most species do show positive correlations with $\rm T_k$. This behaviour positions HC$_3$N as a likely potential ``calibrator'', i.e., a species whose abundance can be used as a reference to compare the relative evolution of other molecules. This would allow more homogeneous comparisons among different kind of regions, especially relevant in evolutionary studies of large samples of interstellar gaseous structures.

The fact that two species with similar functional groups (HCN and HC$_3$N, both containing the cyano radical) behave differently with respect to the temperature, highlights the importance of studying specific chemical pathways. While HCN and its isotopologues appear to reflect material released from dust grains and temperature-dependent chemistry, HC$_3$N seems to be more closely related to gas-phase reactions with lower thermal sensitivity at the studied temperature range (see for instance \citealt{taniguchi2019cyanopolyyne}). 

The results presented here suggest several lines for future research. Extending the analysis to additional similar simple molecular species will help building a more comprehensive picture of the influence of temperature on the chemistry of star-forming cores at the earliest stages. The use of HC$_3$N as a calibrator opens the possibility of establishing normalised abundance ratios, thereby facilitating comparisons across different environments in large-scale observational studies.

\begin{acknowledgement}
R.D.T. is grateful for the support received to attend the 67 Reunión de la Asociación Argentina de Astronomía at Mendoza. This work was partially supported by the Argentinian grants PIP 2021 11220200100012
and PICT 2021-GRF-TII-00061 awarded by CONICET and ANPCYT. 
\end{acknowledgement}


\bibliographystyle{baaa}
\small
\bibliography{ref}

\eject

\end{document}